\documentclass[sigplan,screen]{acmart}

\AtBeginDocument{%
  }

\setcopyright{cc}
\copyrightyear{2026}
\acmYear{2026}
\acmDOI{10.1145/3770743.3804343}
\acmConference[DAC '26]{63rd ACM/IEEE Design Automation Conference}{July 26--29, 2026}{Long Beach, CA, USA}
\acmBooktitle{63rd ACM/IEEE Design Automation Conference (DAC '26), July 26--29, 2026, Long Beach, CA, USA}
\acmISBN{979-8-4007-2254-7/2026/07}

\usepackage{multicol}
\usepackage{multirow}
\usepackage{tabularx}
\usepackage{booktabs} 
\usepackage{amsmath}
\usepackage{subcaption}
\usepackage{xcolor}

\begin{document}

\title{FlowPlace: Flow Matching for Chip Placement}

\author{%
  Peng Xie$^{1,2}$, Ke Xue$^{1,2}$, Yunqi Shi$^{1,2}$, Ruo-Tong Chen$^{1,2,3}$, Chengrui Gao$^{1,2,3}$, \\%
  Siyuan Xu$^{3}$, Chenjian Ding$^{3}$, Mingxuan Yuan$^{3}$, Chao Qian$^{1,2,*}$%
}
\affiliation{%
  \institution{%
    $^{1}$State Key Laboratory of Novel Software Technology, Nanjing University, China\\
    $^{2}$School of Artificial Intelligence, Nanjing University, China\\
    $^{3}$Huawei Noah's Ark Lab, China}
  \country{}}

\renewcommand{\thefootnote}{\fnsymbol{footnote}}
\renewcommand{\shortauthors}{Xie et al.}

\begin{abstract}
Chip placement plays an important role in physical design. While generative models like diffusion models offer promising learning-based solutions, current methods have the following limitations: they use random synthetic data for pre-training, require long sampling times, and often result in overlaps due to their dependence on gradient-based solvers during the sampling process. To overcome these issues, we propose FlowPlace, which features mask-guided synthetic data generation, flow-based efficient training with flexible prior injection, and hard constraint sampling for overlap-free layouts. Experiments on OpenROAD and ICCAD 2015 benchmarks show FlowPlace achieves better PPA metrics, 10-50$\times$ faster sampling efficiency, and zero overlaps.
\end{abstract}

\maketitle

\footnotetext[1]{Corresponding author. E-mail: \texttt{qianc@lamda.nju.edu.cn}.}

\section{Introduction}
In physical design, chip placement is crucial for determining the positions of macros and standard cells, which is typically done in two stages (macro placement and standard cells placement) due to their significant differences in size and quantity~\cite{kahng2023hier,goldie2024chip}. Macro placement forms the foundation for subsequent steps and profoundly influences the final power, performance, and area (PPA)~\cite{wiremask-bbo,kahng2023hier,agnesina2023autodmp}. Thus, automating this complex and critical stage is essential, making reinforcement learning (RL) a promising solution, which typically frames it as a Markov decision process and learns policy through iterative interactions~\cite{nature-graph,deeppr,lai2022maskplace,xue2024reinforcement,geng2024efficient}. However, RL-based placers suffer from poor sample efficiency, weak generalization, and require expensive per-circuit training~\cite{lai2023chipformer}. Their sequential decision-making nature also creates a bottleneck, as early suboptimal choices have irreversible and compounding negative effects.

\begin{figure}[t!]
    \centering

    \setlength{\fboxsep}{0pt}

    \fbox{\includegraphics[width=0.22\columnwidth]{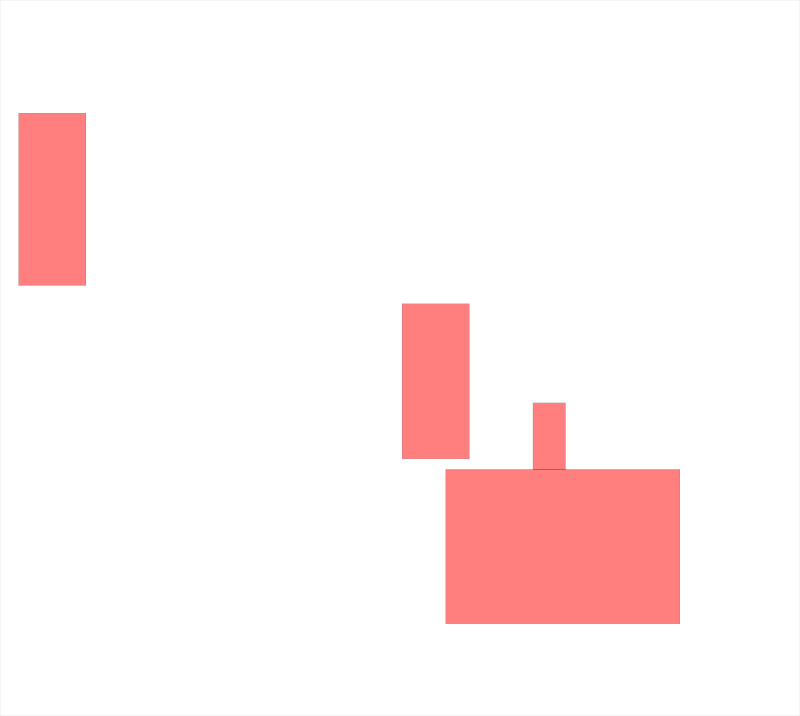}}
    \fbox{\includegraphics[width=0.22\columnwidth]{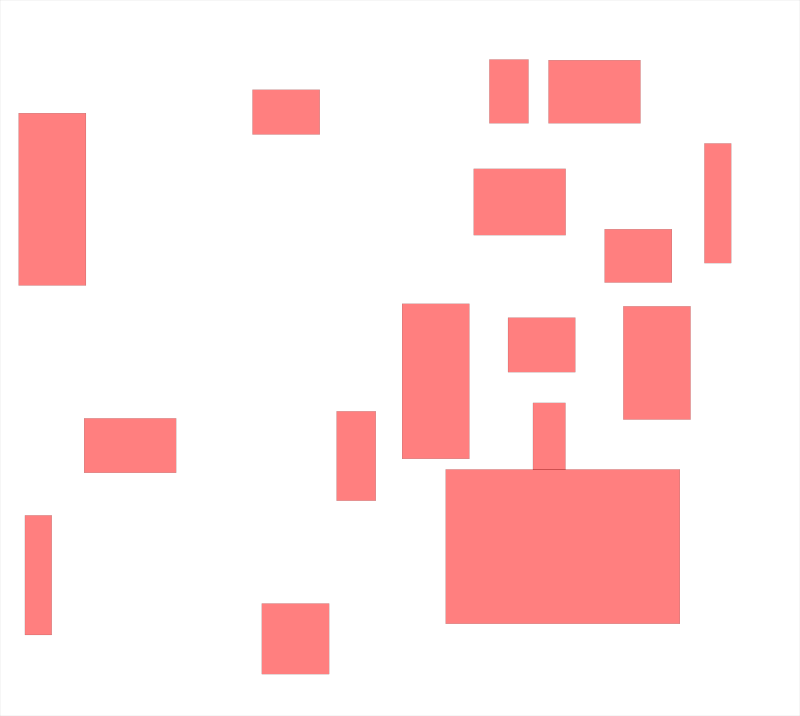}}
    \fbox{\includegraphics[width=0.22\columnwidth]{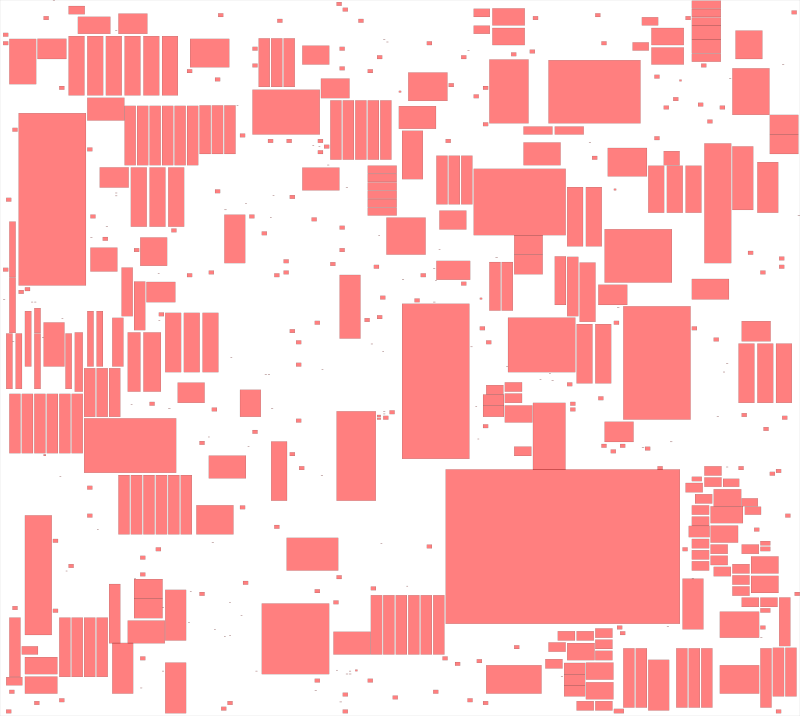}}
    \fbox{\includegraphics[width=0.22\columnwidth]{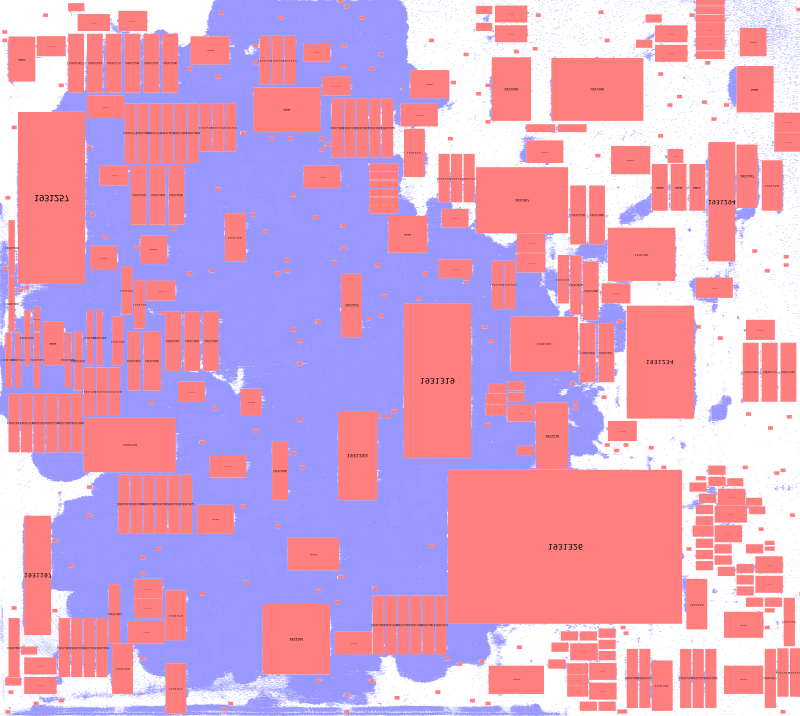}}
    \textbf{EfficientPlace}~\cite{geng2024efficient} \hspace{0.5em} (HPWL = 9.30$\times10^8$,\ \textbf{Overlap = 0$\%$})\\[1pt]

    \fbox{\includegraphics[width=0.22\columnwidth]{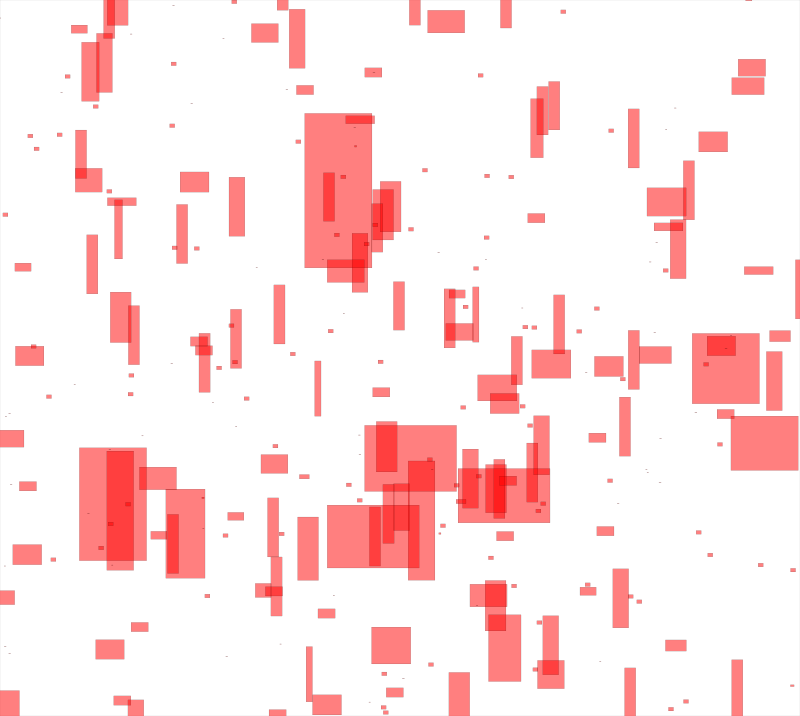}}
    \fbox{\includegraphics[width=0.22\columnwidth]{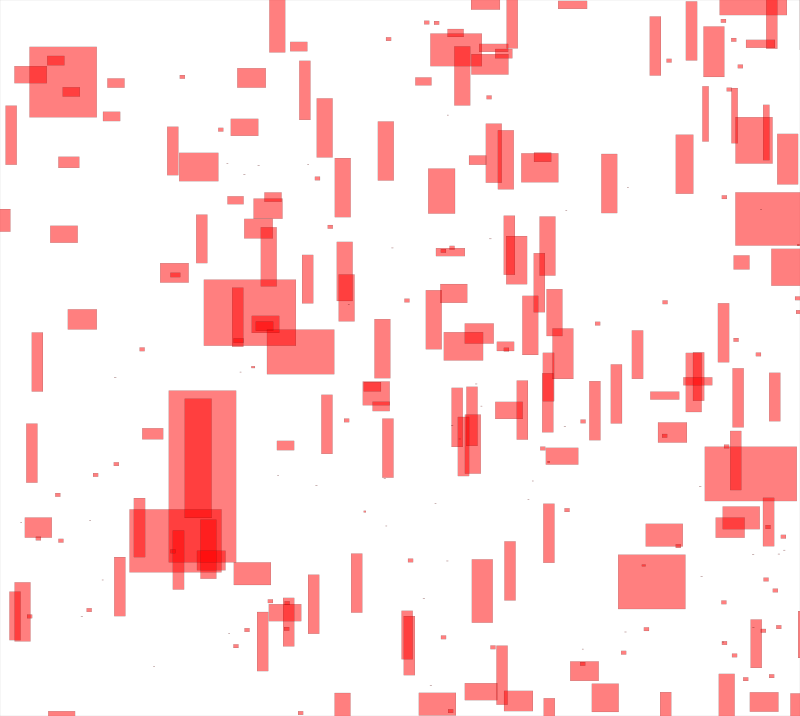}}
    \fbox{\includegraphics[width=0.22\columnwidth]{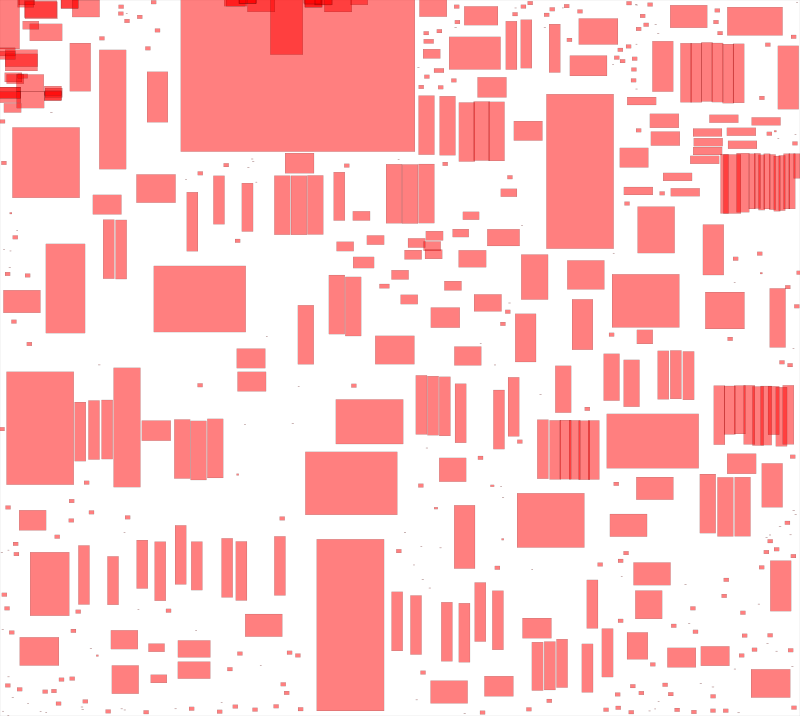}}
    \fbox{\includegraphics[width=0.22\columnwidth]{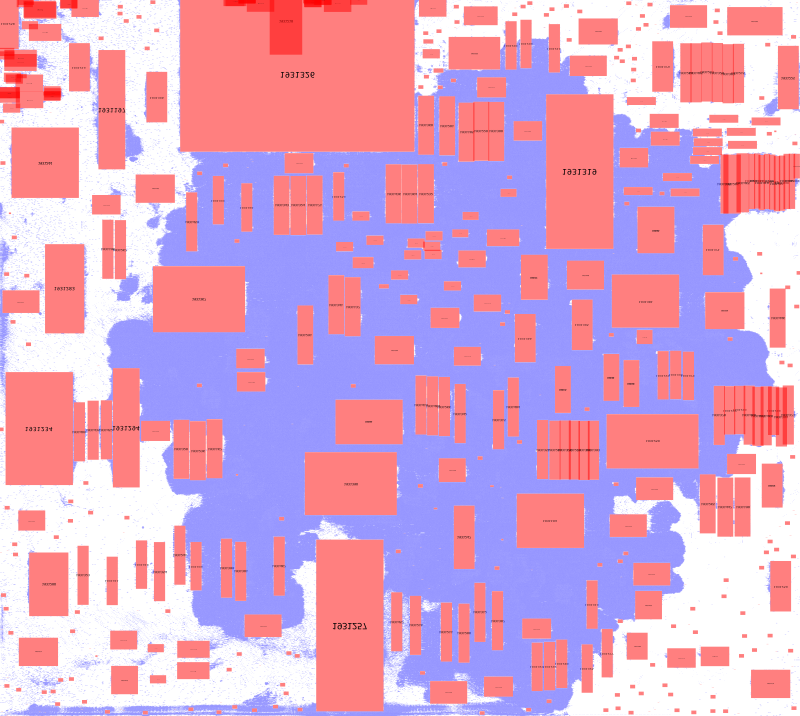}}
    \textbf{ChipDiffusion}~\cite{lee2025chipdiffusion} \hspace{0.5em} (HPWL = 9.56$\times10^8$,\ Overlap = 4.63$\%$)\\[1pt]

    \fbox{\includegraphics[width=0.22\columnwidth]{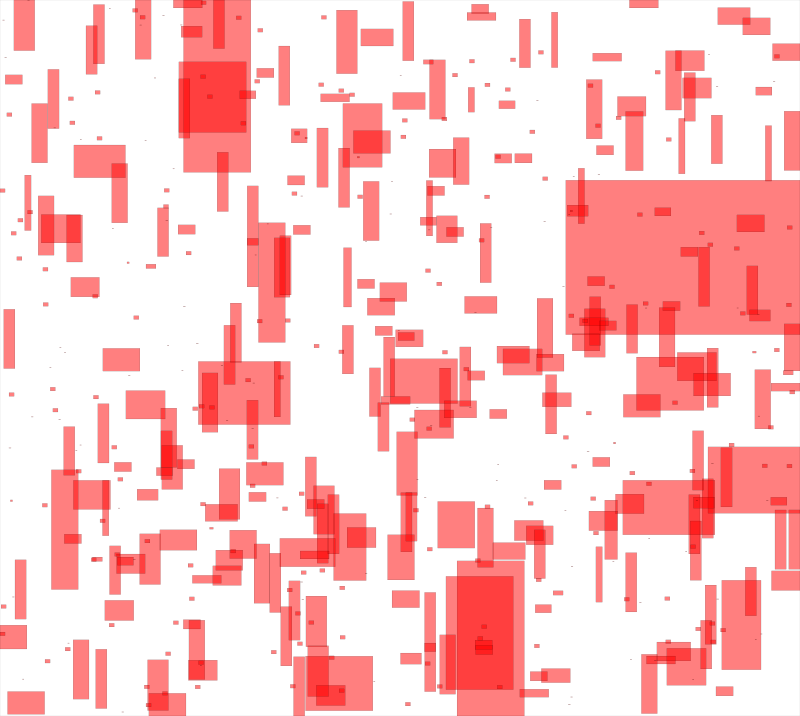}}
    \fbox{\includegraphics[width=0.22\columnwidth]{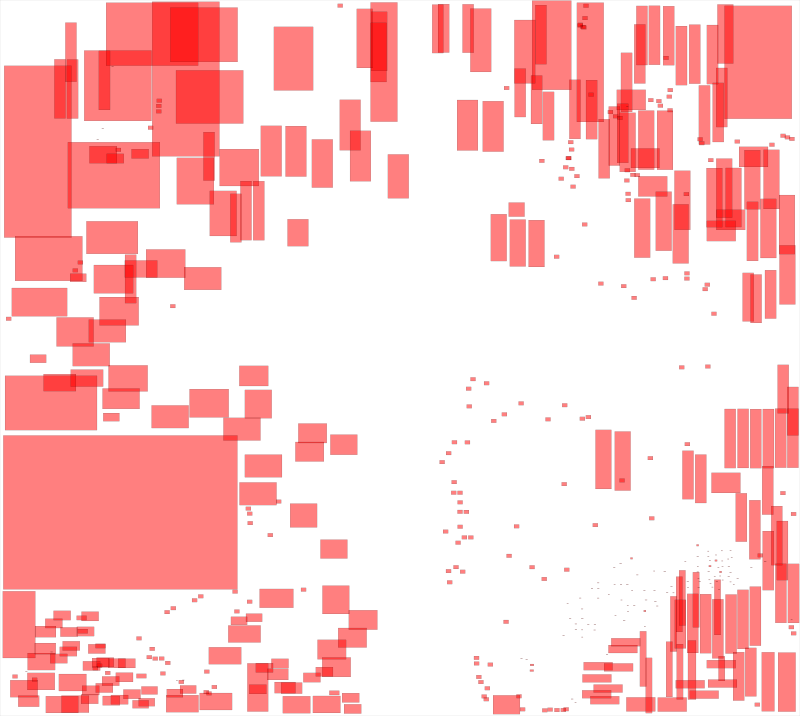}}
    \fbox{\includegraphics[width=0.22\columnwidth]{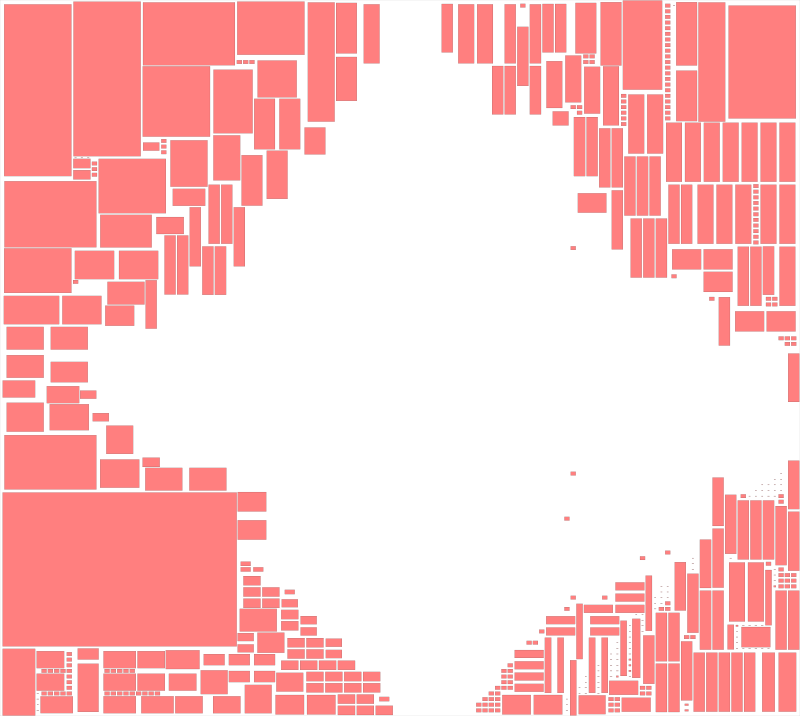}}
    \fbox{\includegraphics[width=0.22\columnwidth]{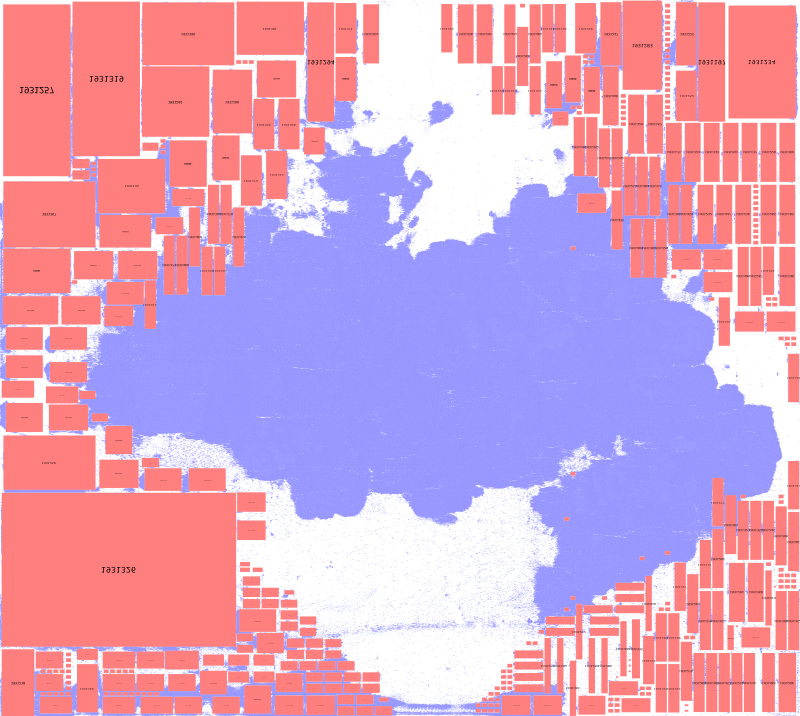}}
    \textbf{FlowPlace (Ours)} \quad (\textbf{HPWL = 8.99$\times10^8$},\ \textbf{Overlap = 0$\%$}) \\
    \vspace{-2mm}
    \caption{Comparison of placement performance on superblue7. EfficientPlace (RL-based) places macros sequentially, while ChipDiffusion and FlowPlace (generative model-based) simultaneously move all macro positions. FlowPlace demonstrates superior performance with the lowest HPWL and zero overlap. }
    \vspace{-4mm}
    \label{fig:trajectory-comparison}
\end{figure}

Recent studies have explored generative learning to address these limitations. ChipDiffusion~\cite{lee2025chipdiffusion} employs a denoising model trained on large-scale synthetic data, then utilizes guided sampling to optimize placement legality and half-perimeter wirelength (HPWL). Unlike RL-based methods that place macros sequentially, this diffusion-based approach places all macros simultaneously in a single step, as illustrated in Figure~\ref{fig:trajectory-comparison}. The diffusion placer achieves zero-shot transfer to unseen real circuits, delivering placements with comparable performance to the RL placer without requiring online fine-tuning.

Despite its potential, the current diffusion-based generative placer exhibits critical weaknesses.  First, although ChipDiffusion is pre-trained on large synthetic datasets, the purely random generation of these datasets lacks domain priors, hindering effective transfer to real-world circuits. Second, diffusion models require long stochastic sampling trajectories, resulting in slow inference and high computational cost. Finally, ChipDiffusion's sampling relies on gradient solvers to handle non-overlap as an additional optimization objective, which is challenging to balance and can still result in illegal placements with constraint violations.

In this paper, we propose a flow matching-based~\cite{lipman2022flow,liu2022flow} generative placer, FlowPlace, to address these critical challenges. 
Firstly, we propose a mask-guided synthetic data generation process that moves beyond random placement. Our approach integrates a modular, boundary-aware sampling prior to generate macro placements with the structural regularities characteristic of high-quality designs, providing high-quality and scalable data for model training. 
Besides, FlowPlace learns a \textit{deterministic} trajectory, instead of learning to reverse a stochastic process via a sequence of denoising steps like diffusion models~\cite{ho2020ddpm,lee2025chipdiffusion}. This offers several advantages, including fewer sampling steps due to deterministic dynamics, flexible choice of prior distribution (rather than the fixed normal Gaussian distribution in diffusion models), and simpler training paradigm without noise schedule tuning.  
Finally, to guarantee non-overlap without relying on inefficient gradient solvers, we introduce a hard constraint guided sampling method. This approach seamlessly integrates a projection operator that directly corrects overlaps into the generative trajectory, progressively steering the layout toward legality while preserving its quality.

In the experiments, FlowPlace is comprehensively evaluated on two established benchmarks: the ICCAD 2015 Contest C benchmark~\cite{iccad15} and the OpenROAD-flow-scripts benchmark suite~\cite{ajayi2019toward}. The results show that FlowPlace achieves superior performance across all key PPA metrics compared to the state-of-the-art analytical placer DREAMPlace 4.1~\cite{dreamplace4}, RL-based placers MaskPlace~\cite{lai2022maskplace} and EfficientPlace~\cite{geng2024efficient}, and the diffusion-based generative method ChipDiffusion~\cite{lee2025chipdiffusion}. A significant advantage of FlowPlace is its ability to generate high-quality layouts in mere seconds through zero-shot inference, as seen in Figure~\ref{fig:percentage-comparison}. This efficiency and effectiveness are attributed to our key designs: a synthetic dataset infused with domain priors and a flexible framework that enables a more suitable prior distribution for the placement task.

This work establishes a foundation for future advanced generative models to tackle increasingly complex design challenges, ultimately paving the way for a fully automated, end-to-end chip placement flow.

\section{Background}
\subsection{Chip Placement Problem}

Chip placement is a critical stage in modern VLSI physical design~\cite{kahng2023hier}, responsible for determining the spatial locations of circuit components. Let $V$ denote the set of movable macros and standard cells, each with size $(w_i, h_i)$, and let $G = (V, E)$ be the netlist hypergraph capturing inter-cell connectivity. The goal is to assign coordinates $\{(x_i,y_i)\}_{i=1}^N$ such that modules do not overlap while optimizing overall chip performance. Because macros often exert a dominant influence on placement quality, we focus on macro placement and defer standard-cell placement to analytical placers.

Classical approaches generally fall into two categories~\cite{shi2025remap}: data structure–based methods and analytical placers. Data structure–based methods~\cite{ousterhout1984corner,murata1996vlsi,hong2000corner,wu2004placement,yan2009handling} use carefully designed representations of placements and apply simulated annealing or evolutionary algorithms for iterative optimization. While these methods strictly enforce feasibility, they suffer from a restricted search space imposed by the data structures and high computational cost, which limits their scalability. Analytical placers~\cite{lin2020dreamplace, chen2023stronger,cheng2018replace, lin2019routability,incre-macro}, in contrast, formulate differentiable objectives for wirelength and density, enabling efficient gradient-based optimization. However, they typically relax or approximate non-overlap constraints and therefore require carefully engineered legalization procedures to ensure manufacturability.

More recently, learning-based methods have gained attention. RL placers~\cite{nature-graph,lai2022maskplace,lai2023chipformer,xue2024reinforcement,geng2024efficient} can capture structural regularities across designs, but they face challenges in training stability and data efficiency~\cite{cheng2023assessment}. 
Diffusion-based generative methods~\cite{lee2025chipdiffusion} further advance automated placement, yet they require complex training pipelines and expensive sampling steps, and often rely on gradient-based correction steps that still do not fundamentally address hard constraint satisfaction (i.e., non-overlap constraint).

\subsection{Flow-based Generative Models}

Let $\mathbb{R}^d$ denote the data space where samples $x_t \in \mathbb{R}^d$ evolve over time $t \in [0,1]$, our goal is to transform a simple base distribution $p_0(x_0)$ into a complex target distribution $p_1(x_1)$ that matches the data distribution $p_{\text{data}}$. Flow-based generative models~\cite{lipman2022flow,liu2022flow} achieve this transformation by defining a time-dependent velocity field $u_t(x): \mathbb{R}^d \times [0,1] \to \mathbb{R}^d$ that induces a continuous flow transporting samples from $p_0$ to $p_1$. Formally, a sample $x_t$ evolves according to the ordinary differential equation (ODE):
$\frac{d}{dt}x_t = u_t(x_t), x_0 \sim p_0(x_0).$
By solving this ODE from $t=0$ to $t=1$, any sample $x_0 \sim p_0$ is transported to $x_1 \sim p_1$, thereby generating samples from the target distribution.

In practice, the true velocity field $u_t$ is unknown and must be learned from data. Flow matching~\cite{lipman2022flow,liu2022flow,tong2024improving}  parameterizes the velocity field with a neural network $v^\theta(x_t, t)$ and trains it using a tractable conditional objective to capture the underlying data distribution $p_{\text{data}}$. The key insight is to construct conditional flows for sample pairs $z = (x_0, x_1)$ where $x_0 \sim p_0$ and $x_1 \sim p_{\text{data}}$. A simple yet effective choice is linear interpolation $x_t = (1-t)x_0 + tx_1$, which induces a constant conditional velocity $v_{t|z}(x_t|z) = x_1 - x_0$. Then, the model can be trained to minimize the conditional flow matching (CFM) loss~\cite{lipman2022flow}:
$\mathcal{L}_{\text{CFM}}(\theta) = \mathbb{E}_{t, z, x_t}\,
\| v^\theta(x_t, t) - v_{t|z}(x_t|z) \|_2^2,$
where $t \sim \mathcal{U}(0,1)$, $z \sim p(z)$, and $x_t \sim p_t(x_t|z)$. Although this objective conditions on sample pairs $z$, it is mathematically equivalent to matching the marginal velocity field $v_t(x_t) = \mathbb{E}_{p(x_t|z)}[v_{t|z}(x_t|z)]$ that generates the marginal probability path $p_t(x_t)$ connecting $p_0$ and $p_1$.

At inference, samples are generated by numerically solving the learned ODE starting from $x_0 \sim p_0$. A typical approach is the Euler method with $N$ discrete steps yields $x_{t+\Delta t} = x_t + \Delta t \cdot v^\theta(x_t, t)$ where $\Delta t = 1/N$, producing samples $x_1$ that approximate the target distribution $p_{\text{data}}$.

\section{FlowPlace}
We introduce the proposed FlowPlace in this section. Our framework integrates three complementary components: (1) Mask-guided synthetic dataset generation to inject domain priors (Section~\ref{sec:3-1}); (2) Flow matching for deterministic, flexible, and efficient training and inference (Section~\ref{sec:3-2}); and (3) Hard constraint guided sampling to ensure non-overlap constraint during generation (Section~\ref{sec:3-3}). The unified framework of FlowPlace achieves faster convergence, higher placement quality, and improved physical design metrics.

\subsection{Mask-guided Synthetic Dataset Generation}
\label{sec:3-1}

Public chip layout datasets are too scarce to train large-scale generative models. ChipDiffusion~\cite{lee2025chipdiffusion} addresses this by treating data synthesis as an \textit{inverse problem}: randomly placing macros on the canvas via rejection sampling to create a layout, then reverse-engineering a netlist by generating pin offsets and connections based on spatial proximity. While this enables large-scale synthetic data creation, the random placement lacks the structural regularities of real designs.

Human designers often place large macros near canvas boundaries~\cite{incre-macro,xue2024reinforcement} to leave central regions for standard cells and routing. This "boundary bias" reduces congestion and improves routability. We adopt the following process to incorporate this domain knowledge prior into our synthetic data generation:

\noindent \textbf{(1) Grid Discretization}: Divide canvas $\mathcal{C}$ into uniform grid cells. Let $\mathcal{P} = \{p_1, \ldots, p_N\}$ denote the set of grid coordinates.

\noindent \textbf{(2) Position Mask Computation}: For each macro $i$ to be placed (ordered by descending size), compute a binary mask $\mathcal{M}_i(p)$ indicating legal positions. This mask computation is parallelized on GPU, avoiding sequential rejection sampling.

\noindent \textbf{(3) Boundary-Aware Scoring}: For each legal location, assign a lightweight heuristic score:
$S(p) = 1/(\text{dist}(p, \partial\mathcal{C}) + \epsilon)^2,$
where $p$ is the legal location coordinate, $\partial\mathcal{C}$ denotes canvas $\mathcal{C}$’s boundary, and $\epsilon > 0$ is a small positive number to prevent division by zero. Higher scores correspond to positions closer to boundaries.
    
\noindent \textbf{(4) Weighted Sampling}: Sample macro $i$'s position from the legal positions with probability proportional to score:
\begin{equation*}
P(p \mid \text{macro } i) = \frac{\mathcal{M}_i(p) \cdot S(p)}{\sum_{p' \in \mathcal{P}} \mathcal{M}_i(p') \cdot S(p')}.
\end{equation*}
\noindent \textbf{(5) Netlist Construction}: Following ChipDiffusion~\cite{lee2025chipdiffusion}, we generate pin offsets on each macro and create connections between spatially proximate pins to form the netlist $G = (V, E)$.

The boundary-bias weight $S(p) \propto 1/\text{dist}^2$ is a simple yet effective heuristic inspired by real placement patterns. Importantly, this scoring mechanism is \textit{modular}—alternative priors can be incorporated by modifying $S(p)$ without changing the overall framework.

\subsection{Flow Matching for Chip Placement}
\label{sec:3-2}

\textbf{Problem Formulation. }We represent the chip placement problem in the flow matching framework. Let $x_t \in \mathbb{R}^{N \times 2}$ denote macro positions at time $t \in [0,1]$, where $N$ is the number of macros. The target distribution $p_1(x_1)$ corresponds to placements from our synthetic dataset, and the base distribution $p_0(x_0)$ is a simple source prior distribution.

\textbf{Model Architecture.} Following ChipDiffusion~\cite{lee2025chipdiffusion} for fair comparison, we represent the netlist $G = (V, E)$ as a graph where node features $\mathbf{h}_i \in \mathbb{R}^2$ encode normalized macro dimensions $(w_i, h_i)$, and edge features $\mathbf{e}_{ij} \in \mathbb{R}^4$ represent pin offset pairs for connections between macros. Our velocity field model $v^\theta(x_t, t, G)$ uses alternating layers of GATv2~\cite{brody2022gatv2} for local connectivity awareness and multi-head self-attention~\cite{vaswani2017attention} for global interactions. Time $t$ is embedded via sinusoidal encoding and fused with node features. The final MLP decoder outputs predicted velocities for all macros.

\textbf{Training.} We adopt linear interpolation between base and target samples: given $x_0 \sim p_0$ and $x_1 \sim p_{\text{data}}$, we construct 
$x_t = (1-t)x_0 + t x_1$ with a constant conditional velocity field 
$v_{t|z}(x_t \mid z) = x_1 - x_0$. 
The model is trained to minimize the conditional flow matching loss~\cite{lipman2022flow}:
$\mathcal{L}(\theta)
= \mathbb{E}_{t, x_0, x_1}\,
\| v^\theta(x_t, t, G) - (x_1 - x_0) \|_2^2$,
where 
$t \sim \mathcal{U}(0,1)$, 
$x_0 \sim p_0$, 
$x_1 \sim p_{\text{data}}$, 
and $x_t = (1-t)x_0 + t x_1$. This direct regression to the constant velocity field enables stable and efficient training without the complex noise scheduling required by diffusion models.

\textbf{Source Distribution.} While diffusion models use fixed Gaussian noise $\mathcal{N}(0, \mathbf{I})$, flow matching allows flexible choice of $p_0$. We note that using a uniform distribution 
over the canvas better matches the geometric properties of the placement space and creates smoother flow trajectories when combined with our boundary-biased target data, consistently improving final metrics.

\textbf{Inference.} At inference, we sample $x_0 \sim p_0$ and solve the ODE $\frac{d}{dt}x_t = v^\theta(x_t, t, G)$ using Euler method with $N$ steps. Intuitively, the trained velocity model $v^\theta$ outputs a velocity vector for each macro at every time step, guiding macros to gradually flow from the initial source distribution $p_0$ toward the target data distribution $p_{\text{data}}$.
The deterministic dynamics of flow matching enable high-quality generation with just 20-50 steps, compared to the requirement of 1000 steps in ChipDiffusion~\cite{lee2025chipdiffusion}.

\subsection{Hard Constraint Guided Sampling}
\label{sec:3-3}

While the learned flow model generates placements efficiently, it does not inherently guarantee hard constraints such as non-overlap. ChipDiffusion~\cite{lee2025chipdiffusion} addresses this via gradient-based guidance, computing analytical gradients $\nabla_{x_t}(\text{HPWL}_{\text{macro}} + \lambda \cdot \text{Overlap})$ at each sampling step to adjust the model's predictions. However, this approach suffers from myopic optimization that focuses on immediate macro-level metrics and may degrade global layout quality. It provides no guarantee of constraint satisfaction as gradient steps may reduce but not eliminate overlaps, and creates strong coupling between generated placement quality and the gradient solver's hyperparameters.

\begin{figure}[t!]
    \centering
    \includegraphics[width=\linewidth]{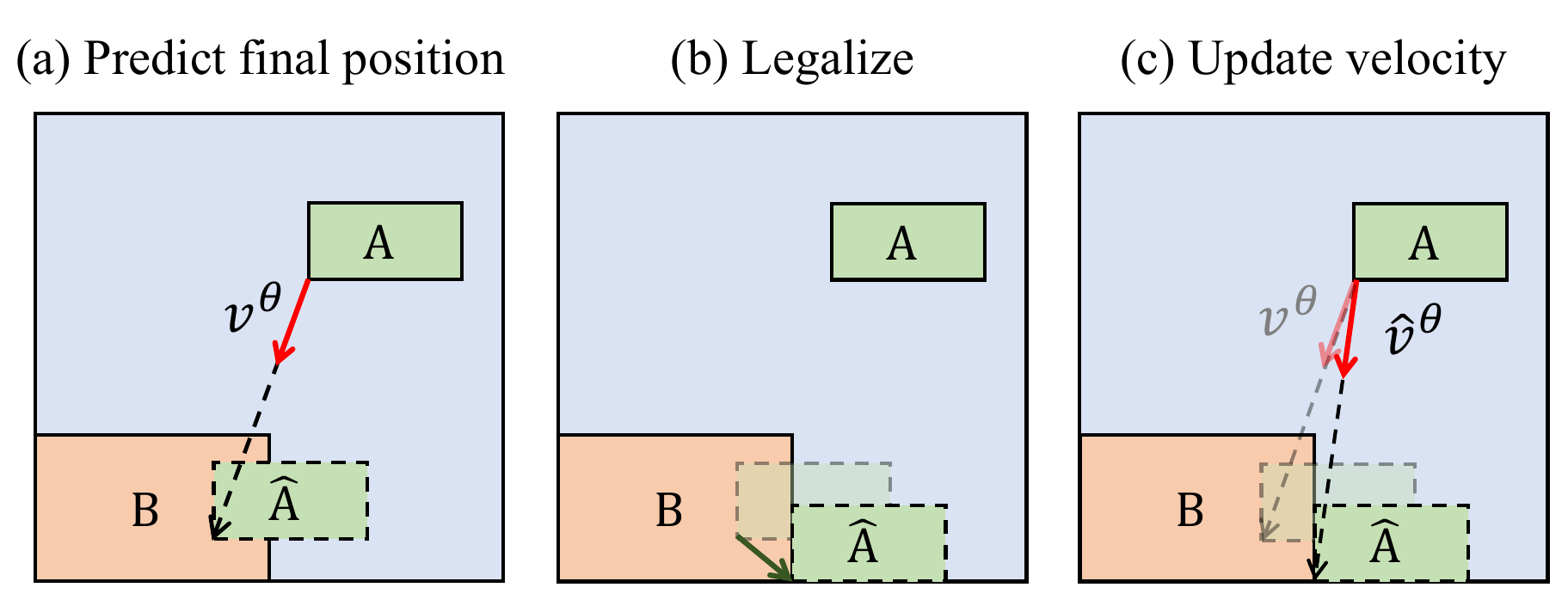}
    \vspace{-4mm}
    \caption{Progressive constraint enforcement process of hard constraint guided sampling. For clarity, we only show one movable macro $A$.}
    \vspace{-4mm}
\label{fig:flow-place} 
\end{figure}

\begin{table*}[t!]
  \centering
  \small
  \renewcommand{\arraystretch}{0.90}
  \caption{Comparison of different methods on ICCAD 2015 benchmark. We report rWL ($\times10^6 \mu$m), rO-H (\%), rO-V (\%), WNS (ns), TNS ($\times10^5$ns), and the respective average ranking. The best and runner-up results per design per metric are highlighted in \textbf{bold} and \underline{underline}, respectively.}
  \vspace{-1em}
  \resizebox{0.9\linewidth}{!}{
      \begin{tabular}{l|l|cccccccc|c}
    \toprule
    Method & Metric & superblue1 & superblue3 & superblue4 & superblue5 & superblue7 & superblue10 & superblue16 & superblue18 & Avg. Rank \\
    \midrule
\multirow{5}{*}{DREAMPlace~\cite{dreamplace4}}
 & rWL & 158.51 & 199.40 & 139.71 & \underline{167.29} & 219.87 & 243.87 & 136.05 & \textbf{53.34} & 4.12 \\ 
 & rO-H & 16.93 & 15.05 & 18.74 & 5.67 & 6.93 & 3.83 & 16.37 & \underline{0.19} & 4.62 \\ 
 & rO-V & 4.72 & 7.40 & 11.60 & 2.18 & 5.24 & 2.10 & 1.93 & \underline{0.02} & 4.62 \\ 
 & WNS & -110.98 & -107.95 & -106.38 & \textbf{-71.72} & \underline{-59.58} & -240.30 & -65.48 & \textbf{-30.26} & 3.50 \\ 
 & TNS & -3.73 & -2.51 & -2.18 & -1.74 & \textbf{-1.37} & -4.10 & \textbf{-2.10} & \textbf{-0.18} & 3.25 \\ 
\midrule
\multirow{5}{*}{MaskPlace~\cite{lai2022maskplace}}
 & rWL & \underline{139.40} & \underline{175.17} & 104.83 & 172.49 & 213.98 & \underline{212.14} & 115.40 & \underline{63.79} & \underline{2.62} \\ 
 & rO-H & 4.61 & 8.91 & 12.00 & 2.74 & 2.41 & 0.64 & 8.82 & 0.70 & 3.50 \\ 
 & rO-V & 0.39 & 0.66 & 0.52 & 0.18 & 0.56 & \textbf{0.05} & 0.16 & 0.14 & 3.38 \\ 
 & WNS & \underline{-77.36} & -101.68 & -71.93 & -169.07 & -92.96 & \textbf{-67.80} & \underline{-50.24} & -35.26 & 3.00 \\ 
 & TNS & -3.15 & -2.27 & -1.69 & -2.21 & -2.44 & -3.52 & -2.55 & -1.04 & 4.25 \\ 
\midrule
\multirow{5}{*}{EfficientPlace~\cite{geng2024efficient}}
 & rWL & 151.24 & 179.52 & \underline{99.38} & 181.77 & \underline{190.40} & \textbf{205.93} & \underline{113.97} & 69.36 & 2.75 \\ 
 & rO-H & \underline{3.73} & \underline{4.70} & \underline{3.96} & 4.07 & \underline{0.39} & \underline{0.23} & \underline{1.09} & 0.56 & \underline{2.38} \\ 
 & rO-V & \underline{0.26} & \underline{0.27} & \underline{0.19} & 0.20 & \underline{0.11} & \underline{0.06} & \underline{0.14} & 0.19 & \underline{2.62} \\ 
 & WNS & -88.42 & \underline{-89.93} & \underline{-69.21} & -111.13 & \textbf{-54.79} & -71.83 & -51.61 & -35.08 & \underline{2.75} \\ 
 & TNS & \underline{-2.73} & -1.89 & \underline{-1.32} & -1.64 & \underline{-1.79} & \underline{-3.42} & \underline{-2.12} & \underline{-0.48} & \textbf{2.25} \\ 
\midrule
\multirow{5}{*}{ChipDiffusion~\cite{lee2025chipdiffusion}}
 & rWL & 152.90 & 179.82 & 109.07 & 189.94 & 206.08 & 218.92 & 123.95 & 69.29 & 3.88 \\ 
 & rO-H & 7.48 & 9.53 & 6.80 & \underline{1.26} & 1.84 & 0.25 & 13.71 & 2.42 & 3.50 \\ 
 & rO-V & 0.36 & 0.28 & 0.30 & \underline{0.12} & 0.25 & 0.22 & 0.18 & 0.12 & 3.12 \\ 
 & WNS & -82.38 & -108.60 & \textbf{-60.04} & \underline{-72.07} & -60.15 & \underline{-71.15} & -58.42 & -36.10 & 3.12 \\ 
 & TNS & -2.90 & \underline{-1.83} & -1.47 & \underline{-1.57} & -2.12 & \textbf{-3.21} & -2.71 & -0.66 & 3.00 \\ 
\midrule
\multirow{5}{*}{FlowPlace}
 & rWL & \textbf{125.77} & \textbf{144.96} & \textbf{86.11} & \textbf{159.25} & \textbf{184.51} & 231.67 & \textbf{107.94} & 65.34 & \textbf{1.62} \\ 
 & rO-H & \textbf{0.29} & \textbf{0.40} & \textbf{0.25} & \textbf{0.05} & \textbf{0.03} & \textbf{0.18} & \textbf{0.65} & \textbf{0.03} & \textbf{1.00} \\ 
 & rO-V & \textbf{0.05} & \textbf{0.06} & \textbf{0.05} & \textbf{0.01} & \textbf{0.04} & 0.12 & \textbf{0.06} & \textbf{0.01} & \textbf{1.25} \\ 
 & WNS & \textbf{-68.71} & \textbf{-88.04} & -74.21 & -80.40 & -95.14 & -82.42 & \textbf{-49.32} & \underline{-34.22} & \textbf{2.62} \\ 
 & TNS & \textbf{-1.86} & \textbf{-1.65} & \textbf{-1.09} & \textbf{-1.33} & -2.06 & -4.17 & -2.45 & -0.55 & \textbf{2.25} \\ 
\bottomrule
  \end{tabular}%
  }
\label{tab:main-results-ppa-innovus}
\end{table*}

To address this issue, we adopt a strategy that progressively enforces constraints~\cite{cheng2025gradientfree} throughout the sampling trajectory. At each time step $t \in \{0, \frac{1}{N}, \frac{2}{N}, \ldots, \frac{N-1}{N}\}$, we perform three operations: (a) We extrapolate using the learned velocity field to predict the final layout $\tilde{x}_1 = x_t + (1-t) \cdot v^\theta(x_t, t, G)$. (b) We apply hard constraint projection $\mathcal{C}(\cdot)$ to obtain a feasible solution $\hat{x}_1 = \mathcal{C}(\tilde{x}_1)$. (c) We compute the corrected velocity $\hat{v}^\theta(x_t, t, G) =\frac{\hat{x}_1 - x_t}{1-t}$ and interpolate the corrected final state back into the trajectory: $x_{t+\Delta t} = (1 - (t+\Delta t)) \cdot x_0 + (t+\Delta t) \cdot \hat{x}_1$. This progressive correction maintains smooth trajectories while steering the generation toward physically valid solutions. The intuitive illustration of the three-step operation at a single time step is provided in Figure~\ref{fig:flow-place}, where each subfigure corresponds to the extrapolation, projection, and corrected velocity-based interpolation steps, respectively. Note that the model $v^\theta$ performs these operations for all macros simultaneously.

To resolve overlaps while preserving layout quality, the constraint projection operator $\mathcal{C}$ performs a greedy legalization. During this process, we select feasible grid cells by jointly considering each macro’s predicted position $\tilde{x}_{1,i}$ and its distance to layout boundaries. The chosen non-overlapping grid location ensures minimal adjustment, and the required geometric checks are efficiently parallelized on GPU, adding negligible overhead to sampling.

\section{Experiments}

\textbf{Benchmarks.} We comprehensively evaluate our method on two benchmark suites. The ICCAD 2015 Contest C Benchmark~\cite{iccad15} comprises 8 large-scale circuits that closely reflect real-world chip design scenarios, with each circuit containing hundreds of macros and millions of standard cells. The OpenROAD-flow-scripts benchmark suite~\cite{ajayi2019toward} consists of 6 real-world designs synthesized using the Nangate45 library. 

\textbf{Evaluation Flow.} For the ICCAD 2015 benchmark, we employ DREAMPlace 4.1.0\footnote{\url{https://github.com/limbo018/DREAMPlace/releases/tag/4.1.0}} to place standard cells, 
and a commercial global router
to execute \texttt{EarlyGlobalRoute} and evaluate PPA metrics including routed wirelength (rWL), routed vertical and horizontal congestion overflow (rO-V, rO-H), worst negative slack (WNS), and total negative slack (TNS). On the OpenROAD-flow-scripts benchmark, we use OpenROAD\footnote{\url{https://github.com/The-OpenROAD-Project/OpenROAD}} to execute the complete design flow including global placement, detailed placement, placement optimization, clock tree synthesis, global routing, and detailed routing, obtaining rWL, WNS, TNS, Power, and Cell Area.

\textbf{Baselines.} We choose DREAMPlace 4.1.0~\cite{chen2023stronger} with two-stage macro placement enhancement and a strong macro legalization technique as the representative analytical placer. For RL-based placers, we compare against state-of-the-art methods including MaskPlace~\cite{lai2022maskplace} and EfficientPlace~\cite{geng2024efficient}, which learn placement policies by sequentially placing macros to maximize cumulative rewards. For generative placers, the diffusion-based ChipDiffusion~\cite{lee2025chipdiffusion} serves as our primary baseline, providing a fair reference for evaluating the effectiveness of our improvements.

\subsection{Main Results}
Table~\ref{tab:main-results-ppa-innovus} presents results on the ICCAD 2015 benchmark~\cite{iccad15}. Our proposed FlowPlace achieves the best or near-best performance across multiple PPA metrics, demonstrating significant improvements over the generative baseline ChipDiffusion~\cite{lee2025chipdiffusion}. 
Compared to RL methods, FlowPlace generates high-quality layouts in just seconds without requiring hours of rollout exploration per circuit. Compared to analytical methods, FlowPlace directly learns the final layout distribution rather than optimizing intermediate proxy metrics, achieving better PPA results even in early design stages.

Table \ref{tab:main-results-ppa-openroad-2f} summarizes results on the OpenROAD benchmark~\cite{ajayi2019toward}. To reflect real design flows, we include full placement-optimization sequences: timing-driven and congestion-driven global placement, plus multiple rounds of buffering and gate sizing both before and after CTS. With only zero-shot inference, FlowPlace consistently achieves the best ranking on all PPA metrics including timing, power, and area. We attribute this to the learned domain priors that yield placements intrinsically amenable to downstream optimization. These results underscore FlowPlace’s generalization capability and practical value in realistic design settings.

\begin{table*}[t!]
  \centering
  \small
  \caption{Comparison of different placement methods on OpenROAD benchmark. We report rWL ($\times10^6 \mu$m), WNS (ns), TNS (ns), Power (W), Cell Area ($\mu$m$^2$), and the respective average ranking.}
  \vspace{-1em}
  \small
  \resizebox{0.85\linewidth}{!}{
    \begin{tabular}{l|l|cccccc|c}
    \toprule
    \textbf{Method} & \textbf{Metric} & \textbf{ariane133} & \textbf{ariane136} & \textbf{bp} & \textbf{bp\_be} & \textbf{bp\_fe} & \textbf{swerv\_wrapper} & \textbf{Avg. Rank} \\
    \midrule
\multirow{5}{*}{DREAMPlace~\cite{dreamplace4}} & rWL   & \textbf{6.86} & \textbf{7.70} & \textbf{9.89} & \underline{2.78} & 2.13 & 5.16 & \textbf{2.17} \\ 
      & WNS   & -1.20 & -1.58 & -4.84 & \textbf{-1.16} & \underline{-0.50} & \underline{-0.82} & \underline{2.67} \\ 
      & TNS   & -3277.38 & -4544.31 & \underline{-31.66} & -353.89 & -115.46 & \underline{-675.28} & 3.00 \\ 
      & Power   & \underline{0.44} & 0.51 & \underline{0.50} & 0.20 & 0.21 & 0.27 & \underline{2.83} \\ 
      & Area   & \textbf{384158} & \textbf{396307} & \textbf{529325} & 123117 & 74222 & 232296 & \underline{2.50} \\ 
\midrule 
\multirow{5}{*}{MaskPlace~\cite{lai2022maskplace}} & rWL   & \underline{7.50} & 8.56 & 10.70 & \textbf{2.56} & 2.25 & 5.08 & 3.50 \\ 
      & WNS   & -1.19 & -1.73 & \textbf{-4.72} & -1.31 & -0.52 & -1.04 & 3.33 \\ 
      & TNS   & -3214.01 & -5068.92 & -668.13 & \underline{-336.85} & -110.83 & -957.62 & 3.67 \\ 
      & Power   & 0.46 & 0.52 & 0.51 & \textbf{0.20} & 0.21 & 0.27 & 3.67 \\ 
      & Area   & 388588 & 402378 & 533643 & \textbf{121154} & 74162 & 230883 & 3.67 \\ 
\midrule 
\multirow{5}{*}{EfficientPlace~\cite{geng2024efficient}} & rWL   & 7.76 & \underline{7.98} & 10.27 & 3.38 & \textbf{2.10} & \underline{4.67} & \underline{2.83} \\ 
      & WNS   & \underline{-1.16} & \textbf{-1.31} & -4.87 & -1.36 & \textbf{-0.49} & -0.95 & 3.00 \\ 
      & TNS   & \underline{-3134.05} & \textbf{-3721.18} & -293.91 & -541.49 & \textbf{-55.03} & -849.27 & \underline{2.67} \\ 
      & Power   & 0.45 & 0.50 & \textbf{0.49} & 0.21 & 0.22 & \underline{0.27} & 3.17 \\ 
      & Area   & 386715 & \underline{397070} & 530593 & 123497 & \underline{74086} & \underline{229884} & 2.83 \\ 
\midrule 
\multirow{5}{*}{ChipDiffusion~\cite{lee2025chipdiffusion}} & rWL   & 8.42 & 8.54 & 10.45 & 3.19 & \underline{2.11} & 4.99 & 3.67 \\ 
      & WNS   & -2.07 & -2.17 & \underline{-4.73} & -1.28 & -0.53 & -0.83 & 3.67 \\ 
      & TNS   & -6789.83 & -10011.40 & -187.28 & -558.75 & \underline{-94.47} & -738.53 & 3.83 \\ 
      & Power   & 0.47 & \textbf{0.50} & 0.52 & 0.21 & \underline{0.21} & 0.27 & 3.33 \\ 
      & Area   & 390776 & 397991 & \underline{530230} & 123317 & 74957 & 230168 & 3.83 \\ 
\midrule 
\multirow{5}{*}{FlowPlace} & rWL   & 7.62 & 8.14 & \underline{10.03} & 3.10 & 2.33 & \textbf{4.54} & \underline{2.83} \\ 
      & WNS   & \textbf{-1.10} & \underline{-1.37} & -4.74 & \underline{-1.19} & -0.56 & \textbf{-0.66} & \textbf{2.33} \\ 
      & TNS   & \textbf{-2874.29} & \underline{-3845.98} & \textbf{-23.87} & \textbf{-316.39} & -124.48 & \textbf{-586.54} & \textbf{1.83} \\ 
      & Power   & \textbf{0.44} & \underline{0.50} & 0.53 & \underline{0.20} & \textbf{0.21} & \textbf{0.27} & \textbf{2.00} \\ 
      & Area   & \underline{384905} & 397798 & 531359 & \underline{121846} & \textbf{72407} & \textbf{227567} & \textbf{2.17} \\ 
\bottomrule
\end{tabular}%
}
  \label{tab:main-results-ppa-openroad-2f}%
\end{table*}

\subsection{Additional Results}\label{sec:4-2}

\begin{figure}[t!]
    \centering
    \begin{minipage}[b]{0.45\linewidth}
        \centering
        \includegraphics[width=\linewidth]{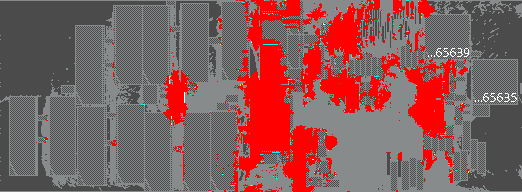}
        \small (a) DREAMPlace~\cite{lin2020dreamplace}
    \end{minipage}
    \hfill
    \begin{minipage}[b]{0.45\linewidth}
        \centering
        \includegraphics[width=\linewidth]{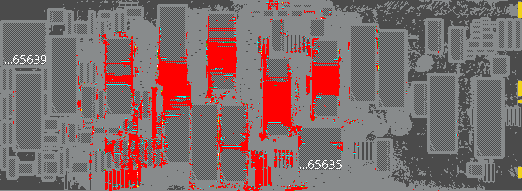}
        \small (b) EfficientPlace~\cite{geng2024efficient}
    \end{minipage}

    \vspace{2pt}

    \begin{minipage}[b]{0.45\linewidth}
        \centering
        \includegraphics[width=\linewidth]{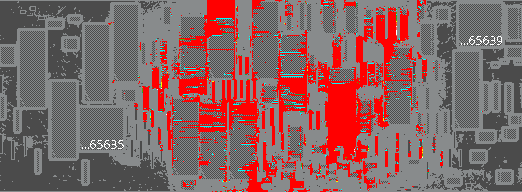}
        \small (c) ChipDiffusion~\cite{lee2025chipdiffusion}
    \end{minipage}
    \hfill
    \begin{minipage}[b]{0.45\linewidth}
        \centering
        \includegraphics[width=\linewidth]{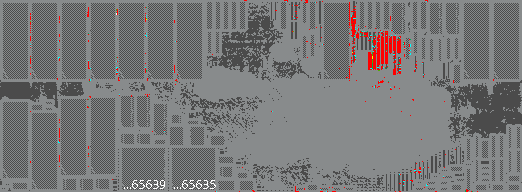}
        \small (d) FlowPlace (Ours)
    \end{minipage}
    \vspace{-1em}
    \caption{Placement layouts and congestion visualization on superblue1. Red points are the congestion critical regions.}
    \label{fig:innovus-congestions-comparison}
    \vspace{-4mm}
\end{figure}

\textbf{Visualization Analysis.} Figure~\ref{fig:innovus-congestions-comparison} compares visualization results from four methods on the ICCAD 2015 benchmark. In particular, Figure~\ref{fig:innovus-congestions-comparison}(d) shows the superblue1 placement produced by FlowPlace: large macros are anchored along the chip boundary, and macros with identical footprints are grouped into tiles, markedly improving placement regularity. This placement leads to substantial reductions in routed congestion, demonstrating that our method effectively learns the injected domain priors and delivers superior PPA outcomes.

\begin{figure}[t!]
    \centering
    \begin{subfigure}[b]{0.47\linewidth}
        \centering
        \includegraphics[width=\linewidth]{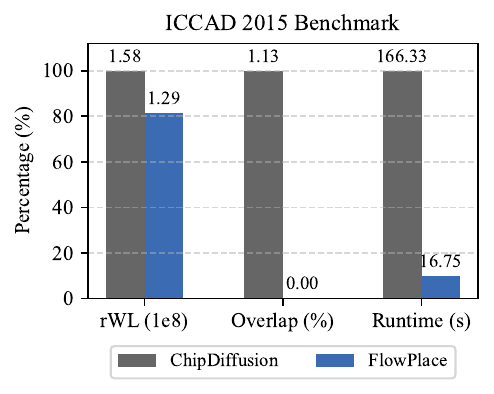}
        \subcaption{ICCAD 2015 benchmark}
        \label{fig:iccad15-percentage}
    \end{subfigure}
    \hfill
    \begin{subfigure}[b]{0.47\linewidth}
        \centering
        \includegraphics[width=\linewidth]{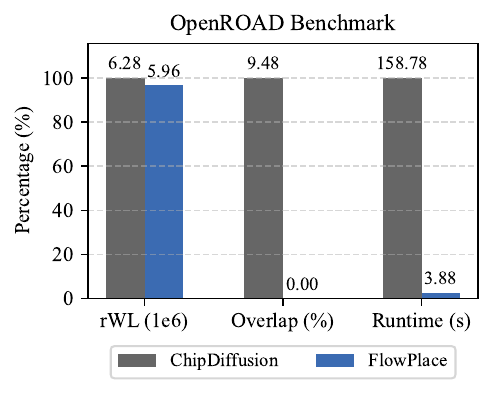}
        \subcaption{OpenROAD benchmark}
        \label{fig:openroad-percentage}
    \end{subfigure}
    \vspace{-2mm}
    \caption{Relative average results of rWL, Overlap, and Runtime between ChipDiffusion and FlowPlace across the two benchmarks. Results for ChipDiffusion are normalized to 100\%, with actual average values annotated above each bar.}
    \vspace{-3mm}
    \label{fig:percentage-comparison} 
\end{figure}

\textbf{Detailed Comparison with The Generative Placer ChipDiffusion.} Figure~\ref{fig:percentage-comparison} compares our method against another generative-model-based placer ChipDiffusion~\cite{lee2025chipdiffusion}.  Our method achieves lower average rWL while guaranteeing zero overlap, and further delivers a 10–50$\times$ speedup. This high efficiency leaves room for subsequent fine-tuning, making FlowPlace more practical for real-world EDA applications.

\begin{table}[t!]
\centering
\small
\caption{Ablation on training datasets: Average PPA metric rank of FlowPlace on the ICCAD 2015 benchmark. Rankings are computed per design across the two training datasets; lower values indicate better performance.}
\vspace{-0.5em}

\resizebox{0.9\linewidth}{!}{
\begin{tabular}{l | c c c c c c}
\toprule
Train Dataset &  rWL  & WNS & TNS & rO-H & rO-V \\
\midrule
Random Synth. & 1.75 & 1.75 & 1.88 & 1.88 & 1.88 \\
\midrule
Mask-Guided Synth. & \textbf{1.25} & \textbf{1.25} & \textbf{1.12} & \textbf{1.12} & \textbf{1.12}  \\
\bottomrule
\end{tabular}
}
\vspace{-1em}
\label{tab:ablation-synthetic-dataset-comparison-ratio}
\end{table}

\begin{table}[t!]
\centering
\small
\caption{Effect of base distributions in FlowPlace: Average HPWL ratio (normalized to the best method per design) on the ICCAD 2015 benchmark. 
The four source priors are standard Gaussian $\mathcal{N}(0,1)$, truncated Gaussian $\mathcal{N}_{[-1,1]}(0,1)$, narrow Gaussian $\mathcal{N}(0,0.5)$, and uniform distribution $U(-1,1)$.}
\vspace{-0.5em}
\resizebox{0.9\linewidth}{!}{
\begin{tabular}{l|c c c c}
\toprule
Source &  $\mathcal{N}(0,1)$ & $\mathcal{N}_{[-1,1]}(0,1)$ & $\mathcal{N}(0,0.5)$ & $U(-1,1)$ \\
\midrule
Average Ratio & 1.067 & 1.023 & 1.029 & \textbf{1.019} \\
\bottomrule
\end{tabular}
}
\label{tab:ablation-source-prior-compact}
\vspace{-1em}
\end{table}

\textbf{Impact of Synthetic Dataset.} Table~\ref{tab:ablation-synthetic-dataset-comparison-ratio} compares our synthetic dataset generated by mask guidance against ChipDiffusion's dataset~\cite{lee2025chipdiffusion} generated randomly. To isolate the effect of the dataset, we train FlowPlace models on each type of data. The result suggests that our proposed domain prior-injected mask-guided dataset consistently achieves superior results across all metrics. Unlike RL or analytical placers constrained by various intermediate proxy objectives, generative methods can directly learn high-quality layout features, achieving higher quality in early design stages and aligning with the shift-left trend in chip design.

\textbf{Choice of Source Prior Distribution.} Table~\ref{tab:ablation-source-prior-compact} analyzes the impact of different base distributions on performance. Unlike ChipDiffusion~\cite{lee2025chipdiffusion} which is fixed to Gaussian priors, FlowPlace's flexible architecture allows us to choose different prior distributions. Leveraging this, we use a uniform prior that better matches the geometric properties of the placement space, leading to improved HPWL quality. Figure~\ref{fig:trajectory-comparison} also shows that compared to ChipDiffusion~\cite{lee2025chipdiffusion}, our uniform prior constrains all macros within the canvas from the start, whereas ChipDiffusion's Gaussian prior places most macros outside boundaries initially, requiring more steps to converge to legal layouts.

\vspace{-0.3em}
\section{Conclusion}
\vspace{-0.3em}

Generative methods present a promising direction for automated chip placement. FlowPlace introduces a flow-based generative framework that incorporates mask-guided generation, deterministic flow matching, and hard-constraint sampling to efficiently produce overlap-free macro placements. This approach directly learns from placement data, circumventing both the local optima of sequential RL methods and the dependency on proxy objectives in analytical placers. As a result, FlowPlace generalizes robustly to unseen circuit designs and achieves superior PPA—even in early optimization stages—while significantly reducing runtime compared to existing methods. The methodology established by FlowPlace opens a path for further development of generative models in tackling more complex chip design problems, contributing to the evolution of end-to-end placement automation.

\vspace{-0.5em}
\section*{Acknowledgment}
This work was supported by the National Science Foundation of China (624B2069), 
the Fundamental Research Funds for the Central Universities (14380020), the Fundamental and Interdisciplinary Disciplines Breakthrough Plan of the Ministry of Education of China (No. JYB2025XDXM118). 

\clearpage
\newpage
\bibliographystyle{ACM-Reference-Format}
\bibliography{dac_reference}

\end{document}